\documentclass[aps,prd,showkeys,showpacs,amssymb,cite,
amsfonts,epsf,preprintnumbers,nofootinbib,superscriptaddress]{revtex4}

\usepackage[dvips]{graphicx}
\usepackage{bm,latexsym,amsmath,amssymb,amsfonts,color}

\newcommand{\be}{\begin{equation}}
\newcommand{\ee}{\end{equation}}
\newcommand{\bear}{\begin{eqnarray}}
\newcommand{\eear}{\end{eqnarray}}
\newcommand{\ba}{\begin{array}}
\newcommand{\ea}{\end{array}}

\begin{document}

\title{Scalar Perturbation Produced at the Pre-inflationary Stage \\
in Eddington-inspired Born-Infeld Gravity}

\author{Inyong Cho}
\email{iycho@seoultech.ac.kr}
\affiliation{Institute of Convergence Fundamental Studies \& School of Liberal Arts,
Seoul National University of Science and Technology, Seoul 139-743, Korea}
\author{Naveen K. Singh}
\email{naveen.nkumars@gmail.com}
\affiliation{Institute of Convergence Fundamental Studies \& School of Liberal Arts,
Seoul National University of Science and Technology, Seoul 139-743, Korea}

\begin{abstract}
We investigate the scalar perturbation produced at the pre-inflationary stage
driven by a massive scalar field in Eddington-inspired Born-Infeld gravity.
The scalar power spectrum exhibits a peculiar rise for low $k$-modes.
The tensor-to-scalar ratio can be significantly lowered
compared with that in the standard chaotic inflation model in general relativity.
This result is very affirmative considering the recent dispute
on the detection of the gravitational wave radiation
between PLANCK and BICEP2.
\end{abstract}
\maketitle

\section{Introduction}
The Eddington-inspired Born-Infeld (EiBI) gravity
is described by the action~\cite{Banados:2010ix},
\begin{eqnarray}\label{action}
S_{{\rm EiBI}}=\frac{1}{\kappa}\int
d^4x\Big[~\sqrt{-|g_{\mu\nu}+\kappa
R_{\mu\nu}(\Gamma)|}-\lambda\sqrt{-|g_{\mu\nu}|}~\Big]+S_{\rm M}(g,\varphi),
\end{eqnarray}
where $\kappa$ is the only additional parameter of the theory
to the gravitational constant $G$ (in this work, we set $8\pi G=1$),
$\lambda$  is a dimensionless parameter,
and $S_{\rm M}(g,\varphi)$ is the action for the matter
which is coupled only to the gravitational field $g_{\mu\nu}$.
The cosmological constant is related by $\Lambda = (\lambda -1)/\kappa$
which we will set to zero in this paper.
The theory follows the Palatini formalism
in which the metric $g_{\mu\nu}$ and the connection
$\Gamma_{\mu\nu}^{\rho}$ are treated as independent fields.
The Ricci tensor $R_{\mu\nu}(\Gamma)$ is evaluated solely by the connection.

The inflationary model with a massive scalar field in this theory
was investigated in Ref.~\cite{Cho:2013pea}.
The matter action is given by
\be \label{S:chaotic}
S_{\rm M}(g,\varphi) = \int d^4 x \sqrt{-|g_{\mu\nu}|}
\left[ -\frac12 g_{\mu\nu} \partial^\mu\varphi \partial^\nu \varphi -V(\varphi) \right],
\qquad
V(\varphi) = \frac{m^2}{2} \varphi^2,
\ee
which is the same form as for the chaotic inflation model \cite{Linde:1983gd}
in general relativity (GR).

Due to the square-root type of the EiBI action,
there is an upper bound in pressure beyond which the theory is not defined.
In the maximal pressure state (MPS),
the scale factor exhibits an exponential expansion of the Universe.
The MPS is the past attractor
from which all the evolution paths of the Universe originate \cite{Cho:2013pea}.
At early times, the energy density is very high at the MPS
(the magnitudes of the field $\varphi$ and its velocity are very large).
The Hubble parameter becomes $H_{\rm MPS} \approx 2m/3$,
and thus the curvature scale remains constant.
In describing the high-energy state of the early universe,
therefore, quantum gravity is not necessary .

The MPS is known to be unstable under the global perturbation (zero-mode scalar perturbation) \cite{Cho:2013pea}.
With a small perturbation, the Universe evolves to the near-MPS stage,
and then enters into the intermediate stage which is followed by an inflationary attractor stage.
The inflationary feature at the attractor stage was found
to be the same as the ordinary chaotic inflation in GR \cite{Cho:2013pea}.

As a whole, there are two inflationary stages in the EiBI inflation model,
the near-MPS stage and the attractor stage.
(Since the inflationary feature at the attractor stage is similar to the standard inflation,
we call the near-MPS stage as the pre-inflationary stage for convenience.)
If the initial conditions drive the Universe to evolve in such a way
to acquire the sufficient 60 $e$-foldings at the attractor stage,
the cosmological situation must be very similar to that of the standard chaotic inflation.
However, if they do not, the supplementary $e$-foldings should be provided
at the near MPS-stage to solve cosmological problems.
In this case, the story of the density perturbation may be altered in the long-wavelength modes.

The tensor perturbation in the EiBI inflation model was investigated in Ref.~\cite{Cho:2014ija}.
For short-wavelength modes, the perturbation is very similar to that of the standard
chaotic inflation in GR, with a small EiBI correction.
The tensor power spectrum is smaller than that in GR
as $P_{\rm T} \approx P_{\rm T}^{\rm GR}/(1+\kappa m^2\varphi_i^2/2)$,
where $\varphi_i$ is the value of the scalar field at the beginning of the inflationary attractor stage.
For long-wavelength modes, there exists a peculiar rise in the power spectrum $P_{\rm T}$
originated from the near-MPS stage.

The scalar perturbation produced at the inflationary attractor stage
was investigated in Ref.~\cite{Cho:2014jta} in the limit of $\kappa m^2 \ll 1$.
The scalar power spectrum is smaller than that in GR
as $P_{\cal R}^{\rm ATT} \approx (1-4\kappa m^2/3) P_{\cal R}^{\rm GR}$.
From the results of the tensor and the scalar spectra,
the tensor-to-scalar ratio $r$ becomes smaller than that in GR
as $r \approx [(1+4\kappa m^2/3)/(1+\kappa m^2\varphi_i^2/2)] r^{\rm GR}$.

In this paper, we investigate the scalar perturbation produced at the near-MPS stage
in the very early universe.
One question is whether or not the peculiar phenomenon arises for long-wave modes
as in the tensor mode, which may leave a signature in the cosmic microwave background radiation.
The other is how much different the tensor-to-scalar ratio would be
from that for the perturbation produced at the attractor stage.
The results are summarized as following.
The peculiar rise in the scalar power spectrum $P_{\cal R}$ for long-wave length modes
appears exactly in the same way as the tensor spectrum $P_{\rm T}$ in EiBI theory.
The ratio $r$ is not different from that for the perturbation
produced at the attractor stage
because the EiBI correction turns out to be exactly the same.

The density perturbation in EiBI gravity has been investigated
in Refs.~\cite{Lagos:2013aua,EscamillaRivera:2012vz,Avelino:2012ue,Yang:2013hsa}
for the Universe with the perfect-fluid background.
Some other works on cosmology and astrophysics in EiBI gravity can be found in
Refs.~\cite{Cho:2012vg,Pani:2011mg,Pani:2012qb,DeFelice:2012hq,Avelino:2012ge,Avelino:2012qe,Casanellas:2011kf,
Liu:2012rc,Delsate:2012ky,Pani:2012qd,Cho:2013usa,Scargill:2012kg,Kim:2013nna,Kim:2013noa,Du:2014jka,Ji:2014hna,
Fu:2014raa,Harko:2014nya}.

This paper is organized as following.
In Sec. 2, we present the background field equations and the perturbation equations in the literature
following Refs. \cite{Banados:2010ix,Cho:2014ija,Cho:2014jta,Lagos:2013aua}.
In Sec. 3, we apply the near-MPS approximation on the equations,
and obtain the solution for the scalar perturbation.
In Sec. 4, following the solution-matching technique used in Ref. \cite{Cho:2014ija},
we derive the power spectrum and obtain the tensor-to-scalar ratio.
In Sec. 5, we conclude.

\section{Field equations}
In this section, we present the background field equations,
the scalar perturbations and their field equations
introduced in Ref.~\cite{Cho:2014jta}.

\subsection{Background Field Equations}
When there is no cosmological constant ($\lambda =1$),
the EiBI action \eqref{action} is equivalent to a bimetric-like theory action
\be\label{S}
S[g,q,\varphi] = \frac{1}{2} \int d^4x \sqrt{-|q_{\mu\nu}|} \left[  R(q) - \frac{2}{\kappa}  \right]
+{1 \over 2\kappa} \int d^4x \left(\sqrt{-|q_{\mu\nu}|}q^{\alpha\beta}g_{\alpha\beta} - 2\sqrt{-|g_{\mu\nu}|}
\right)+ S_{\rm M}[g,\varphi],
\ee
where $g_{\mu\nu}$ is the physical metric and $q_{\mu\nu}$ is the auxiliary metric
by which the affine connection $\Gamma$ is determined.
From this action, the equations of motion are obtained as
\begin{align}
\frac{\sqrt{-|q|}}{\sqrt{-|g|}}~q^{\mu\nu}
& =\lambda g^{\mu\nu} -\kappa T^{\mu\nu},\label{eom1}\\
q_{\mu\nu} & = g_{\mu\nu}+\kappa R_{\mu\nu}, \label{eom2}
\end{align}
where $T^{\mu\nu}$ is the energy-momentum tensor in the standard form.
The ans\"atze for the metrics are
\begin{align}
q_{\mu\nu}dx^\mu dx^\nu &= b^2(\eta) \left[ -\frac{d\eta^2}{z(\eta)} +\delta_{ij}dx^idx^j \right],\\
g_{\mu\nu}dx^\mu dx^\nu
&= -dt^2 +a^2(t) \delta_{ij}dx^idx^j
= a^2(\eta) \left( -d\eta^2 +\delta_{ij}dx^idx^j \right),
\end{align}
where $t$ and $\eta$ are the cosmological and the conformal time, respectively.
For the derivatives, we denote as
$\hat{} \equiv d/dt$, $' \equiv \d/d\eta$,
${\cal H} \equiv a'/a$, $H \equiv \hat{a}/a$, and $h\equiv b'/b$.
The components of Eq.~\eqref{eom1} give
\be\label{eom1cp}
b^2\sqrt{z}=\left(1+ \kappa\rho_0\right)a^2,
\qquad
\frac{b^2 }{\sqrt{z}}=\left(1 - \kappa p_0\right)a^2,
\ee
where the subscript $0$ stands for the unperturbed background fields, i.e.,
$\rho_0 = \varphi_0'^2/2a^2 + V(\varphi_0)$ and $p_0 = \varphi_0'^2/2a^2 - V(\varphi_0)$.
From Eq.~\eqref{eom1cp}, we get $z=(1+\kappa\rho_0)/(1 - \kappa p_0)$.
The dynamical equations for the metric coefficients
are obtained from the components of Eq.~\eqref{eom2} as
\begin{align}
b^2 &= 3\kappa z \left(\frac{b'}{b}\right)^2 -\frac{a^2}{2} (z-3), \label{eqn_b1}\\
b^2 &= a^2 +\kappa z \left[ \frac{b''}{b} + \left(\frac{b'}{b}\right)^2
+ \frac{1}{2}\frac{b'}{b}\frac{z'}{z}\right], \label{eqn_b2}
\end{align}
and the scalar field equation is in the standard form,
\be\label{Seq}
\varphi_0''+2{\cal H}\varphi_0' +a^2 \frac{dV}{d\varphi_0} =0.
\ee
We can get the background solutions $a$, $b$, $z$, and $\varphi_0$ by solving
Eqs. \eqref{eom1cp}-\eqref{Seq}.

\subsection{Scalar Perturbations and Their Equations}
Let us introduce the scalar perturbations for $q_{\mu\nu}$ and $g_{\mu\nu}$ as
\begin{align}
ds_q^2 &= b^2\left\{-\frac{1+ 2\phi_1}{z} d\eta^2  +2\frac{B_{1,i}}{\sqrt{z}} d\eta dx^i
+ \Big[(1-2\psi_1)\delta_{ij} + 2 E_{1,ij}\Big] dx^i dx^j \right\}, \label{qpert} \\
ds_g^2 &= a^2\left\{-(1+ 2\phi_2) d\eta^2 + 2 B_{2,i}d\eta dx^i
+\Big[(1-2\psi_2)\delta_{ij} + 2 E_{2,ij}\Big]dx^i dx^j \right\},  \label{metric_pert}
\end{align}
and the scalar-field perturbation is given by $\varphi=\varphi_0 +\chi$.
Plugging these perturbed metrics and the scalar field into the action \eqref{S},
we can write the second-order action for the perturbation fields as $S_{\rm s}=S_1+S_2+S_3$,
where
$S_1$ involves the perturbation fields for $q_{\mu\nu}$,
$S_2$ involves the perturbation fields for $g_{\mu\nu}$ and the mixing terms with $q_{\mu\nu}$,
and $S_3$ involves the matter-field perturbation,
as was presented in Ref. \cite{Cho:2014jta},
\begin{align}\label{S1}
S_{1}[\phi_1,B_1,\psi_1,E_1]
&= \frac{1}{2}\int d^4x\; \Bigg\{  \frac{b^2}{\sqrt{z}} \left[4zh\psi_1' E_{1,ii}-6z\psi_1^{'2} \right.
-12zh(\phi_1+\psi_1)\psi'_1-2\psi_{1,i}(2\phi_{1,i}-\psi_{1,i} ) \nonumber \\
&- 4h\psi_{1,i}B_{1,i} +6zh^2(\phi_1+\psi_1)E_{1,ii}-4\sqrt{z}h(\phi_1+\psi_1)(B_1-\sqrt{z}E_1')_{,ii} \nonumber \\
&- 4\sqrt{z}\psi_1'(B_1-\sqrt{z}E_1')_{,ii}
-4\sqrt{z}h E_{1,ii}(B_1-\sqrt{z} E'_1)_{,jj}
 +4\sqrt{z}hE_{1,ii}B_{1,jj} \nonumber \\
&+ 3zh^2E_{1,ii}E_{1,jj} + 3zh^2B_{1,i}B_{1,i} \left.-9zh^2(\phi_1+\psi_1)^2\frac{}{}\right] \nonumber \\
&- \frac{2 b^4}{\kappa\sqrt{z}}\left[\frac{3}{2}\psi_1^2-3\phi_1\psi_1+\frac{1}{2}B_{1,i}B_{1,i}\right.
 \left.-\frac{1}{2}E_{1,ii}E_{1,jj}-\frac{1}{2}\phi_1^2+E_{1,ii}(\phi_1-\psi_1)\right] \Bigg\},
\end{align}
\begin{align}\label{S3}
S_{2}[\phi_k,B_k,\psi_k,E_k]
&= \frac{1}{2}\int d^4x \; \Bigg\{ \frac{a^2 b^2}{\kappa\sqrt{z}} \Big[2\sqrt{z}B_{1,i}B_{2,i}
+ \phi_1\left[\left(z-1\right)\left(3\psi_1-E_{1,ii}\right)-6\psi_2  +2E_{2,ii}-2z\phi_2\right] \nonumber \\
&+ \psi_1 \left[6\psi_2-(z-1)E_{1,ii}-2E_{2,ii}-6z\phi_2\right] - \frac{1}{2}(z-1)(E_{1,ii}E_{1,jj}+B_{1,i}B_{1,i})  \nonumber \\
&+ \frac{3}{2}\left(\phi_1^2+\psi_1^2\right)(z-1) -  2E_{1,ii}\left(\psi_2-z\phi_2+E_{2,ii}\right)\Big]
 \nonumber \\
&-\frac{2a^4}{\kappa}\left[\frac{3}{2}\psi^2_2-\frac{1}{2}\phi_2^2
+ \frac{1}{2}B_{2,i}B_{2,i} -\frac{1}{2}E_{2,ii}E_{2,jj}+(\phi_2-\psi_2)E_{2,ii} -3\phi_2\psi_2\right] \Bigg\},
\end{align}
\begin{align}\label{S3}
S_{3}[\phi_2,B_2,\psi_2,E_2,\chi]
&= \frac{1}{2} \int d^4x\; a^2\Bigg\{ \varphi_0'^2 \left(4 \phi_2^2 - B_{2,i}B_{2,i} \right) \nonumber \\
&+ \left(\varphi_0'^2 - 2 V_0 a^2\right) \left[ \frac{1}{2} \left( 3 \psi_2^2 - \phi_2^2
+ B_{2, i} B_{2, i}- E_{2,ii}E_{2,ii}\right) - 3 \phi_2 \psi_2 + (\phi_2-\psi_2) E_{2,ii}\right]\nonumber \\
&-2 \varphi_0' \chi_{, i} B_{2,i} - 4 \varphi_0' \chi' \phi_2 + \chi'^2
+ 2 \left(\phi_2-3\psi_2 + E_{2, ii}\right) (\chi' \varphi_0' - V_1 a^2 - \phi_2 \varphi_0'^2) -\chi_{, i}\chi_{, i} - 2V_2 a^2 \Bigg\}.
\end{align}
Here, $V_i$ is the $i$th-order potential
from $V=V_0(\varphi_0)+V_1(\chi)+V_2(\chi)$.
Using $\rho_0$ and $p_0$, $S_3$ can be rewritten as
\begin{align}
S_{3}[\phi_2,B_2,\psi_2,E_2,\chi]
&= \int d^4 x a^4 \Bigg\{
p_0 \left[ \frac{1}{2} \left( 3 \psi_2^2 - \phi_2^2+ B_{2, i} B_{2, i}- E_{2,ii}E_{2,ii}\right)
- 3 \phi_2 \psi_2 + (\phi_2-\psi_2) E_{2,ii}\right]  \nonumber \\
&+ (\rho_0 +p_0) \left[ 2\phi_2(\phi_2 -{\cal X}\chi') -\frac{1}{2}B_{2,i}(B_{2,i} +2{\cal X}\chi_{,i})
+(\phi_2-3\psi_2 + E_{2, ii})({\cal X}\chi' -{\cal Y}\chi - \phi_2)   \right] \nonumber \\
&+ \frac{1}{2 a^2}(\chi'^2-\chi_{,i} \chi_{,i}) - \frac{m^2}{2} \chi^2\Bigg\},
\end{align}
where
\be\label{XY}
{\cal X} \equiv \frac{1}{a\sqrt{\rho_0+p_0}},\qquad
{\cal Y} \equiv -m\frac{\sqrt{\rho_0-p_0}}{\rho_0+p_0}.
\ee
(Please note that there was a typo in the definition for ${\cal Y}$ in  Ref. \cite{Cho:2014jta};
the signature ``$-$" was missing.)

Denoting nine perturbation fields as $F_l$ ($l=1\sim 9$),
we introduce the Fourier modes as
\be\label{Fourier}
F_l(\eta,\vec{x}) = \int \frac{d^3k}{(2\pi)^{3/2}}
F_l(\eta,\vec{k})e^{i\vec{k}\cdot\vec{x}}.
\ee
The gauge conditions for the scalar-field type and the perfect-fluid type
have been rigorously studied in Ref.~\cite{Lagos:2013aua}.
For the scalar-field type,
one element of two sets given below can be fixed,
\bear
\left(\psi_1, \psi_2, \chi\right) + \left(E_1, E_2\right).
\eear
We fix the gauge conditions as
\be\label{gg}
\psi_1=0 \quad{\rm and}\quad E_1=0.
\ee
Performing the variation for $S_2$ and $S_3$ with respect to $\phi_2$, $\psi_2$, $E_2$, and $B_2$,
we get
\begin{align}
& (1-2z)\phi_2 +z\phi_1 -3z\psi_2 -k^2zE_2 -(1-z){\cal X}\chi' -(1-z){\cal Y}\chi =0, \label{eqn_phi2} \\
& 3\psi_2 + 3\phi_1 - 3z\phi_2 +k^2E_2 - 3(1-z){\cal X}\chi' +3(1-z){\cal Y}\chi =0, \label{eqn_psi2}\\
& k^2E_2 - \phi_1 + z\phi_2 - \psi_2 +(1-z){\cal X}\chi' -(1-z){\cal Y}\chi =0, \label{eqn_E2}\\
& zB_2 - \sqrt{z}B_1 -(1-z){\cal X}\chi =0, \label{eqn_B2}
\end{align}
and for $S_1$ and $S_2$ with respect to $\phi_1$ and $B_1$,
we get
\begin{align}
& (6\kappa h^2 -a^2)z\phi_1 +a^2z\phi_2 +3a^2\psi_2 +k^2a^2E_2 -2k^2\kappa h\sqrt{z}B_1 =0, \label{eqn_phi1} \\
& a^2B_1 -2\kappa h\sqrt{z}\phi_1 - a^2\sqrt{z}B_2  =0. \label{eqn_B1}
\end{align}
From Eqs. \eqref{eqn_psi2} and \eqref{eqn_E2}, we get $E_2 = 0$.
From Eqs. \eqref{eqn_phi2} and \eqref{eqn_E2}, we get
\be
\phi_2 = \frac{(z-1)(3z+1){\cal X}\chi' -(z-1)(3z-1){\cal Y}\chi +4z\phi_1}{(z+1)(3z-1)} \label{int_phi2},
\ee
and from Eqs. \eqref{eqn_B2} and \eqref{eqn_B1}, we get
\be
\phi_1 = \frac{a^2(z-1){\cal X}\chi}{2 \kappa hz}.
\ee
Then we finally get from Eqs. \eqref{eqn_psi2} and \eqref{int_phi2},
\be\label{psi2_XY}
\psi_2 = \frac{z-1}{2\kappa hz(z+1)(3z-1)}
\Big[ -2\kappa hz(z-1){\cal X}\chi' +a^2(z-1)^2{\cal X}\chi +2\kappa hz(3z-1){\cal Y}\chi \Big],
\ee
which is expressed only by the matter-field perturbation $\chi$ and the background fields.
This result for $\psi_2$ will be used in evaluating the comoving curvature later.

Using the results of Eqs.~\eqref{eqn_phi2}-\eqref{psi2_XY},
$S_{\rm s}[\chi]$ is expressed only
by $\chi$ and the background fields in the Fourier space as
\be\label{Ss}
S_{\rm s}[\chi] = \frac{1}{2}\int  d^3k d\eta\;   \Big[ f_1(\eta,k) \chi'^{2} - f_2(\eta,k)  \chi^2 \Big],
\ee
where
\be\label{f1}
f_1(\eta,k) = a^2 +\frac{2a^2(z-1)^2{\cal X}^2\left[ a^2(z-3)-6\kappa h^2z \right]}{\kappa\sqrt{z}(z+1)(3z-1)},
\ee
and
\be\label{f2}
f_2(\eta,k) = \frac {\beta}{8\kappa^3 h^2 z^{5/2}(z + 1)^2}.
\ee
Here,
\be
\beta= a^2 \left[\frac{\beta_1}{3z-1} + \frac{\beta_2}{(3z-1)^2}\right],
\ee
where
\begin{align}
\beta_1 &= (z+1)\Bigg\{ 8\kappa^3h^2z^2(3z-1)\Big[ k^2\sqrt{z} - 12 h^2{\cal Y}^2 z
+ k^2 z^{3/2} + 24 h^2 {\cal Y}^2 z^2 - 12 h^2 {\cal Y}^2 z^3 - 3 k^2 h^2  {\cal X}^2 (z-1)^2(z+1)\Big]  \nonumber \\
&+ a^6 {\cal X}^2 (z-3)(z-1)^3(3z^2-2z +3)
+ 4 \kappa a^4 h {\cal X} z(z-1)^2 \Big[ {\cal Y}(z-3)^2(3z-1) - 3 h {\cal X} z(3z^2 -6z -1)\Big] \nonumber \\
 &+ 4 \kappa^2 a^2 h^2 z(3z-1) \Big[-6h {\cal X}{\cal Y}(z-3)(z-1)^2 z + {\cal X}^2(z-1)^2(z+1)[(k^2+ 9h^2)z -3k^2] \nonumber \\
 &+ 4{\cal Y}^2z(z-3)(z-1)^2   + 2\kappa m^2 z^{3/2}(z+1) \Big]\Bigg\}, \\
\beta_2 &= (z-1) \Big[a^2 ( z-3) - 6 \kappa h^2  z\Big]
\Bigg\{ a^4 {\cal X}^2 (z-1)^2(z+1)(3z-1)(3z^2-2z+3)\nonumber \\
&+ 4 \kappa^2 h^2 z(3z-1)^2 \Big[2z(z-1)(z+1) [ 2(h+\mathcal{H}){\cal X}{\cal Y} +\left({\cal X}{\cal Y}\right)']
+ {\cal X}{\cal Y}(z^2 + 6z + 1) z' \Big] \nonumber \\
&+ 2 \kappa a^2 h {\cal X} \Big[z(z-1)(z+1)(3z-1)(3z^2-2z+3) [(h + 4{\cal H}){\cal X} + 2{\cal X}']  \nonumber \\
&+ {\cal X}(9 z^5 + 21z^4 - 34 z^3 + 30 z^2 + 9 z  -3)z'\Big] \Bigg\}. \label{beta2}
\end{align}

We can construct a perturbation field $Q$ in the canonical form from the action \eqref{Ss}
by a transformation $Q=\omega\chi$
with introducing a new time coordinate $\tau$ by $d\tau =(\omega^2/f_1)d\eta$.
The field equation then becomes
\be\label{Q-eq1}
\ddot{Q} + \left(\sigma_s^2 k^2-\frac{\ddot{\omega}}{\omega}\right)Q =0 ,
\ee
where $\dot{} \equiv d/d\tau$ and $\sigma_s^2 \equiv f_1 f_2/k^2\omega^4$.
We consider a Bunch-Davies vacuum in the $k\to\infty$ limit.
Requiring $\sigma_s^2 \to 1$ in this limit determines $\omega$ to be
\begin{align}\label{omega}
\omega^4 &= \frac{a^4}{2 \kappa^2 z^2(z+1)(3z-1)} \Bigg\{ a^2 {\cal X}^2 (z-3)(z-1)^2
- 2 \kappa z \Big[ 3 h^2 {\cal X}^2 (z-1)^2 -\sqrt{z}\Big]\Bigg\} \nonumber \\
 &\times \Bigg\{ 2 a^2 {\cal X}^2 (z-3)(z-1)^2
 - \kappa \sqrt{z}\Big[12 h^2 {\cal X}^2 \sqrt{z} (z-1)^2  -3 z^2 -2 z +1 \Big]\Bigg\}.
\end{align}
The normalization condition for the canonical field is given by
\be\label{norm}
Q\dot Q^*-Q^*\dot Q =i.
\ee

\section{Production of Perturbation at near-MPS stage}
In this section, we investigate the scalar perturbation produced at the near-MPS stage
by solving the perturbation equation \eqref{Q-eq1}.
In order to do that, we need to know $f_1$, $f_2$, and $\omega$ as a function of $\tau$.
First, we evaluate them in terms of $a$ and $z$,
and then get the expressions in $\tau$.

$f_1(a,z)$ and $\omega(a,z)$ are obtained in general,
while $f_2(a,z)$ is obtained only in the near-MPS approximation.
Let us get $f_1(a,z)$ and $\omega(a,z)$ first.
The equation \eqref{eqn_b1} can be rewritten as
\be\label{h}
h^2 = \frac{1}{3\kappa z}\left[ b^2 +\frac{1}{2}a^2(z-3) \right].
\ee
Using Eq.~\eqref{eom1cp}, we get
\be\label{eqn_Xnew}
{\cal X}^2 = \frac{1}{a^2\left(\rho_0+p_0\right)} = \frac{\kappa \sqrt{z}}{b^2(z-1)}.
\ee
Using these two equations \eqref{h} and \eqref{eqn_Xnew},
$f_1$ and $\omega$ in Eqs. \eqref{f1} and \eqref{omega} can be simplified as
\be
f_1(a,z) = \frac{a^2 (3z^2-2z+3)}{(z+1)(3z-1)},
\ee
and
\be\label{omega4}
\omega^4(a,z) = \frac{a^4 (3 z^2-2 z+3)}{z(z+1)(3z-1)}.
\ee

\subsection{Perturbation Equation in Near-MPS Approximation}
In this subsection, we introduce the near-MPS solutions investigated in Ref. \cite{Cho:2013pea},
and apply the approximation using them in order to get $f_2(a,z)$.
Then we get $a(\tau)$ and $\omega(\tau)$ in this approximation,
and finally get the perturbation equation in terms of $\tau$.

The maximal pressure state (MPS) is achieved when $p_0=1/\kappa$,
\be\label{mpc}
\frac{1}{\kappa} - p_0 =\frac{1}{\kappa} - \frac{\hat{\varphi}_0^2}{2} + V(\varphi_0) = 0.
\ee
When $p_0>1/\kappa$, the action \eqref{action} becomes imaginary,
and the theory is not well defined.
As it was studied in Ref. \cite{Cho:2013pea}, however,
the MPS state is the past attractor of all the evolution paths of the Universe.
If we flush back in time, the Universe takes infinite time to reach this state.
The path never crosses the MPS,
and the ill-defined region of the pressure is dynamically inaccessible.
The MPS solution for the background scalar field is obtained from this condition \cite{Cho:2013pea},
\be\label{mps_sol_phi}
\varphi_0(t) = \sqrt{\frac{2}{\kappa m^2}} \sinh(mt),
\ee
where we considered the scalar field initially rolling down the potential
in the region of $\varphi_0<0$ (so $\dot\varphi_0 >0$).
At the MPS, the Friedmann equation reduces to $H=\hat a/a = -(2/3) dU/d\varphi_0$,
where $U \equiv \sqrt{2 [1/\kappa + V(\varphi_0)]} = \sqrt{2/\kappa + m^2\varphi_0^2}$,
and the scale factor is solved as \cite{Cho:2013pea}
\be\label{mps_sol_a}
a(t) = a_0 U^{-2/3} = a_0 \left(\frac{\kappa}{2} \right)^{1/3} \cosh^{-2/3}(mt) ,
\ee
where $a_0$ is an integration constant.
The MPS was found to be unstable under the global perturbation in Ref.~\cite{Cho:2013pea}.
With a small perturbation, the Universe leaves the MPS and evolves to the near-MPS.
Introducing small perturbations $\psi(t)$ and $\gamma(t)$
for the velocities of the scalar field and the scale factor
at the near-MPS as
\begin{align}
\hat{\varphi}_0 &= U \Big[1+ \psi(t)\Big], \label{deriphi} \\
H &= \frac{\hat{a}}{a} = -\frac{2}{3} \frac{d U}{d \varphi_0} \Big[1+ \gamma(t)\Big],
\end{align}
the solutions were found as \cite{Cho:2013pea}
\begin{align}\label{psi}
\psi &= \psi_0 U^{-4/3} e^{t/t_c},\\
\gamma &= \psi_0 \left(-\frac{2}{3} + \sqrt{\frac{2}{3\kappa}} \frac{d\varphi_0}{dU} \right) U^{-4/3} e^{t/t_c},
\end{align}
where $t_c = \sqrt{3 \kappa/8}$.
Then at the near-MPS, using the above results, we have
\begin{align}
\frac{1}{\kappa}-p_0 &= -\psi \left(1+\frac{1}{2}\psi\right) U^2 , \label{lambda-p}\\
\frac{1}{\kappa} +\rho_0 &= \left( 1+ \psi +\frac{1}{2}\psi^2\right) U^2, \label{lambda+rho}
\end{align}
which give
\be\label{psieqn}
z= \frac{1+ \kappa\rho_0}{1- \kappa p_0}
= -\frac{1}{\psi}\left( \frac{1+ \psi + \psi^2/2}{1 + \psi/2}\right) \gg 1
\quad\Longrightarrow\quad
\psi = -1 +\sqrt{\frac{z-1}{z+1}}.
\ee
Therefore, for the near-MPS approximation, we assume $z \gg 1$.
From Eqs. \eqref{eom1cp} and \eqref{lambda-p}-\eqref{psieqn},
we have
\be\label{b4}
b^4 = (1+\kappa\rho_0)(1-\kappa p_0) a^4 = \frac{\kappa^2 a^4 U^4 z}{(z+1)^2}.
\ee
Using Eq. \eqref{b4}, $h$ in Eq. \eqref{h} can be expressed as
\be\label{eqnhb}
h^2 = \frac{a^2(z-3)}{6\kappa z} \left[1+ \frac{2 \kappa U^2 \sqrt{z}}{(z+1)(z-3)}\right],
\ee
and ${\cal X}^2$ in Eq. \eqref{eqn_Xnew} can be approximated as
\be\label{eqnX2}
{\cal X}^2 = \frac{1}{a^2 U^2}\frac{(z+1)}{(z-1)}
\approx \frac{1}{a^2 U^2} \approx \frac{a}{a_0^3}
\qquad\Rightarrow\qquad
\frac{{\cal X}'}{{\cal X}} \approx \frac{{\cal H}}{2},
\ee
where we used the relation $a=a_0 U^{-2/3}$ in Eq. \eqref{mps_sol_a}.
Using Eqs. \eqref{mps_sol_phi} and \eqref{mps_sol_a}, we get
\be\label{eqnY2}
{\cal Y} = -m\frac{\sqrt{\rho_0-p_0}}{\rho_0+p_0}
=m^2\frac{\varphi_0}{\hat{\varphi}_0^2}
=m \left( \frac{a}{a_0} \right)^{3/2} \tanh(mt)
\approx -m \left( \frac{a}{a_0} \right)^{3/2}
\qquad\Rightarrow\qquad
\frac{{\cal Y}'}{{\cal Y}} \approx \frac{3}{2}{\cal H}.
\ee
Differentiating Eq.~\eqref{eqn_b1} with respect to $\eta$ and using Eq.~\eqref{eqn_b2},
we get
\be\label{eqn_dz}
\frac{z'}{z}= - 4\frac{b'}{b} + 2\frac{a'}{a}\left(\frac{3}{z}-1\right)
= -4h +2{\cal H}\left(\frac{3}{z}-1\right)
\approx -4h -2{\cal H} .
\ee

Plugging the approximations \eqref{eqnX2}-\eqref{eqn_dz} into Eqs. \eqref{f2}-\eqref{beta2},
and keeping the $k$-dependent term and the highest-order terms in $z$ ($z^0$ terms),
we get
\be\label{f2app1}
f_2 \approx a^2 \left( \frac{k^2}{z} + m^2a^2 \right)
- 4a^2 \left( \frac{\cal Y}{\cal X} \right)^2
- a^4 \left( 6\frac{\cal H}{a^2} +\frac{1}{\kappa h} \right) \frac{\cal Y}{\cal X}
-\frac{a^6}{4\kappa^2 h^2}
-\frac{3a^4}{2\kappa} \left( \frac{\cal H}{h} -1 \right).
\ee
Using
${\cal Y}/{\cal X} \approx -ma$ from Eqs. \eqref{eqnX2} and \eqref{eqnY2},
${\cal H}=aH \approx 2ma/3$ from Eq. \eqref{mps_sol_a},
and $h \approx a/\sqrt{6\kappa}$ from Eq. \eqref{eqnhb},
one can show that the last four terms in Eq. \eqref{f2app1} cancel,
and $f_2$ can be further approximated as
\be
f_2(a,z) \approx a^2 \left( \frac{k^2}{z} + m^2a^2 \right).
\ee
Using the results of $f_1(a,z)$, $f_2(a,z)$, and $\omega(a,z)$,
the perturbation equation \eqref{Q-eq1} is approximated at the near-MPS stage as
\be\label{Q-eqaz}
\sigma_s^2k^2 = \frac{f_1f_2}{\omega^4} \approx k^2 + m^2a^2z
\quad\Rightarrow\quad
\ddot{Q} + \left( k^2 + m^2a^2z -\frac{\ddot{\omega}}{\omega}\right)Q \approx 0.
\ee

Now let us express this perturbation equation in terms of $\tau$.
From Eqs. \eqref{mps_sol_a}, \eqref{psieqn}, and \eqref{psi},
at the near-MPS stage, we have
\be\label{atzt}
a(t) \approx a_0(2\kappa)^{1/3} e^{2mt/3},
\qquad
z(t) \approx -\frac{1}{\psi}
\approx -\frac{1}{\psi_0 (2\kappa)^{2/3}}e^{-(\sqrt{8/3\kappa}+4m/3)t}.
\ee
The time coordinates are transformed for $z\gg1$ as
\be
d\tau =\frac{\omega^2}{f_1}d\eta = \frac{\omega^2}{af_1}dt
\approx \frac{dt}{a\sqrt{z}}
\approx \frac{\sqrt{-\psi_0}}{a_0} e^{\sqrt{2/3\kappa}t} dt
\quad\Rightarrow\quad
\tau \approx \sqrt{-\frac{3\kappa\psi_0}{2a_0^2}} e^{\sqrt{2/3\kappa}t},
\ee
where we set $\tau =0$ for $t\to-\infty$.
In terms of $\tau$, we have
\begin{align}
a(\tau) &\approx a_0(2\kappa)^{1/3}
\left( -\frac{2a_0^2}{3\kappa\psi_0}\right)^{\sqrt{\kappa m^2/6}} \tau^{\sqrt{2\kappa m^2/3}},\label{atau}\\
z(\tau) &\approx -\frac{1}{\psi_0(2\kappa)^{2/3}}
\left( -\frac{2a_0^2}{3\kappa\psi_0} \right)^{-(1+\sqrt{2\kappa m^2/3})}
\tau^{-2(1+\sqrt{2\kappa m^2/3})}.\label{ztau}
\end{align}
With these results of $a(\tau)$ and $z(\tau)$,
we get
\be
\sigma_s^2k^2 \approx  k^2 + m^2a^2z \approx k^2 + \frac{3\kappa m^2}{2\tau^2},
\ee
and from Eq. \eqref{omega4} for $z\gg1$, we get
\be
\omega^4 \approx \frac{a^4}{z}
\qquad\Rightarrow\qquad
\frac{\ddot\omega}{\omega} \approx \left( -\frac{1}{4} +\frac{3}{2}\kappa m^2\right) \frac{1}{\tau^2}.
\ee
Finally, the perturbation equation \eqref{Q-eqaz} at the near-MPS stage becomes
\be\label{Q-eq2}
\ddot{Q} + \left( k^2 + \frac{1}{4\tau^2} \right)Q \approx 0.
\ee
The $\kappa$-dependent EiBI corrections in $\sigma_s^2k^2$ and $\ddot\omega/\omega$
cancel each other, and the resulting equation is in the same form
with the near-MPS equation for the tensor perturbation
studied in Ref.~\cite{Cho:2014ija}.

\subsection{Near-MPS Solution}
Since the perturbation equation \eqref{Q-eq2} is in the same form with that for the tensor perturbation
at the near-MPS stage, we manipulate the solution here in the same way as in Ref. \cite{Cho:2014ija}.
The solution to Eq. \eqref{Q-eq2} is given by
\be\label{Qsol}
Q(\tau) = \sqrt{\tau}\Big[ c_1J_0(k\tau) + c_2Y_0(k\tau) \Big].
\ee
Here, $c_1$ and $c_1$ are complex,
\be\label{c1c2}
c_1 = c_1^{\rm Re} + ic_1^{\rm Im} \equiv c,
\qquad
c_2 = c_2^{\rm Re} + ic_2^{\rm Im} \equiv R-i \frac{\pi}{4c},
\ee
where $c$ and $R$ are real.
One arbitrariness was fixed by imposing $c_1^{\rm Im}=0$,
and $c_2^{\rm Im}$ was determined from by normalization condition \eqref{norm}.
$R$ and $c$ are to be determined by imposing the initial condition
at the moment of the production of the perturbation.
The initial condition is given by minimizing the energy,
\be\label{EE}
E =\frac12 \left[|\dot Q|^2 + \left(k^2 +\frac1{4\tau^2} \right)|Q|^2\right].
\ee
The production moment $\tau_*$ of the perturbation is determined as following.
As it was studied in Ref.~\cite{Cho:2013pea},
the curvature scale $H$ is finite at the beginning of the Universe
at $\tau =0$ ($t\to -\infty$ and $\varphi \to -\infty$),
so the quantum gravity is not necessary.
However, the wavelength scale of the perturbation becomes smaller than the Planck scale.
To treat the perturbation in a classical way,
we consider the production of the perturbation
when the wavelength scale $\lambda_{\rm phys}$ is comparable to the Planck scale $l_{p}$,
\be\label{lplanck}
\lambda_{\rm phys} = \frac{a(\tau_*)}{k} \gtrsim l_p
\qquad\Rightarrow\qquad
\tau_* \gtrsim a^{-1} (kl_p)
\approx \sqrt{-\frac{3\kappa\psi_0}{2a_0^2}}
\left[ \frac{kl_p}{a_0(2\kappa)^{1/3}} \right]^{\sqrt{3/2\kappa m^2}}.
\ee

For high $k$-modes, the perturbation is produced
after the solution $Q$ in Eq. \eqref{Qsol}
is relaxed to the oscillatory behavior ($k\tau_* \gg 1$).
Using the asymptotic formulae for $k\tau \gg 1$,
it was found in Ref.~\cite{Cho:2014ija} that
the energy $E$ in Eq. \eqref{EE} is minimized when $R=0$ and $c^2=\pi/4$.
Then the perturbation solution for high $k$-modes becomes
\be\label{QMPS_highk}
Q(\tau) = \pm \frac{1}{\sqrt{2k}}e^{i\pi/4}e^{-ik\tau},
\ee
which is the plane-wave solution with only the positive energy mode selected.

For low $k$-modes, the perturbation is produced
before the solution $Q$ is relaxed to the oscillatory behavior ($k\tau_* < 1$).
(Please see Fig. 1 in Ref. \cite{Cho:2014ija}.)
In this case, the energy $E$ in Eq. \eqref{EE} is minimized when
\be\label{cR}
c^2 = \frac{\pi}4 \frac{Y^2 +Y_0^2}{|JY_0-J_0Y|},
\qquad
R=\mp \sqrt{\frac{\pi}{4}}\frac{JY + J_0 Y_0}{\sqrt{|JY_0-J_0Y|(Y^2 +Y_0^2)}},
\ee
where $J \equiv {(J_0-2k\tau_* J_1)}/{\sqrt{1+ 4k^2\tau_*^2}}$,
$Y \equiv {(Y_0-2k\tau_* Y_1)}/{\sqrt{1+ 4k^2\tau_*^2}}$,
$J_{0,1} \equiv J_{0,1}(k\tau_*)$, and $Y_{0,1} \equiv Y_{0,1}(k\tau_*)$.
(Please see Ref. \cite{Cho:2014ija} for detailed calculations.)
With these $c$ and $R$, the solution \eqref{Qsol} for low $k$-modes becomes
\be\label{QMPS_lowk}
Q(\tau)  = \sqrt{\tau}\left[c J_0(k\tau)
+ \left( R-i\frac{\pi}{4c}\right)Y_0(k\tau) \right] .
\ee

As a whole, the near-MPS solutions were obtained
as Eq. \eqref{QMPS_highk} for high-$k$ modes,
and as Eq. \eqref{QMPS_lowk} low-$k$ modes.
The coefficients were fixed by imposing the minimum-energy condition
at the production moment $\tau_*$.
These two perturbation modes produced at the near-MPS stage
evolve to the {\it intermediate stage} which is connected to the inflationary {\it attractor stage}.

\section{Power Spectrum}
The power spectrum $P_{\cal R}$ is evaluated at the end of the inflationary attractor stage,
while the coefficients of the mode solution at the attractor stage
are determined from the initial perturbation produced at the near-MPS stage.
In order to determine the coefficients of the mode solution $Q_{\rm ATT}$ at the attractor stage
from the near-MPS solution $Q_{\rm MPS}$,
we assume that the perturbation evolves adiabatically from the near-MPS stage
through the intermediate stage till the attractor stage.
The adiabatic period spans from the late near-MPS stage to the early attractor stage,
and is described by the WKB solution $Q_{\rm WKB}$.
In order to determine the coefficients,
we need the solution matching between $Q_{\rm MPS}$ and $Q_{\rm WKB}$
at some moment $\tau_1$ at the late near-MPS stage,
and between $Q_{\rm WKB}$ and $Q_{\rm ATT}$ at some moment $\tau_2$
at the early attractor stage.

The tensor perturbations at all the stages were investigated in Ref. \cite{Cho:2014ija}.
In Ref. \cite{Cho:2014jta}, the scalar perturbation at the attractor stage was investigated,
and the solution $Q_{\rm ATT}$ was found.
Both of $Q_{\rm ATT}$ and $Q_{\rm MPS}$ (obtained in this paper)
are exactly in the same form respectively with those of the tensor perturbation
obtained in Ref. \cite{Cho:2014ija}.
In this section, therefore, we shall follow the solution-matching technique
exactly in the same manner as for the tensor perturbation.
(For details, please see Ref. \cite{Cho:2014ija}.)
We shall focus on the low $k$-modes, then the result is applied for the high $k$-modes
simply by setting $R=0$ and $c^2=\pi/4$ in the end.

The near-MPS solution is
\be
Q_{\rm MPS}(\tau) = \sqrt{\tau}\Big[ c_1J_0(k\tau) + c_2Y_0(k\tau) \Big],
\ee
where $c_1$ and $c_2$ are given by Eqs. \eqref{c1c2} and \eqref{cR}.
In the adiabatic period, the solution to the perturbation equation
$\ddot Q + \Omega_k^2(\tau) Q =0$ is given by the WKB approximation,
\be\label{sol:WKB}
Q_{\rm WKB}(\tau) = \frac{b_1}{\sqrt{2\Omega_k(\tau)}}
    \exp\left[ i \int^\tau \Omega_k(\tau') d\tau'\right]
+ \frac{b_2}{\sqrt{2\Omega_k(\tau)}}
    \exp\left[ -i \int^\tau \Omega_k(\tau') d\tau'\right],
\ee
for which the adiabatic condition is
$\Omega_k^{-3}\left| d\Omega_k^2/d\tau \right| \ll 1$.
The attractor solution was obtained in Ref. \cite{Cho:2014jta},
\begin{align}
Q_{\rm ATT}(\tau)  &= A_1 \left[ \cos k(\tau-\tau_0) - \frac{\sin k(\tau-\tau_0)}{k(\tau-\tau_0)} \right]
+A_2 \left[ \sin k(\tau-\tau_0) + \frac{\cos k(\tau-\tau_0)}{k(\tau-\tau_0)} \right] \nonumber \\
&= A_1' \left[1+ \frac{i}{k (\tau-\tau_0)}\right] e^{ik(\tau-\tau_0)}
+A_2'  \left[1- \frac{i}{k (\tau-\tau_0)}\right] e^{-ik(\tau-\tau_0)},\label{muATTmt2}
\end{align}
where $A_1'= (A_1-i A_2)/2$, $A_2' =(A_1+i A_2)/2$,
$\tau_0 \equiv \tau_i - \sqrt{6}/\varphi_im a_i$,
and the subscript $i$ stands for the values at the beginning of the attractor stage.

Now we match $Q$'s and $\dot Q$'s at $\tau_1$ for MPS and WKB
and at $\tau_2$ for WKB and ATT.
Then from the results in Ref. \cite{Cho:2014ija},
the coefficients are determined as
\begin{align}
b_{1,2} &\approx \frac{c_1 \mp i c_2}{\sqrt{\pi}} \,e^{ \pm i (k\tau_1 -\pi/4)},\label{b12}\\
A_{1,2}' &\approx \frac{e^{\mp ik(\tau_2-\tau_0)}}{2}
\left[ Q_{\rm WKB}(\tau_2;b_1,b_2) \mp \frac{i}{k} \dot Q_{\rm WKB}(\tau_2;b_1,b_2) \right]. \label{A12}
\end{align}
At the end of inflation, the perturbation is approximated as
$Q_{\rm ATT}(\tau) \approx i(A_1'-A_2')/k(\tau-\tau_0)$,
and with the aid of above equations \eqref{b12} and \eqref{A12},
one can get
\be
|Q_{\rm ATT}|^2 \approx
\frac{|A_1'-A_2'|^2}{k^2(\tau-\tau_0)^2}
=  \frac{c^2+R^2 + \pi^2/16c^2}{\pi k^3 (\tau-\tau_0)^2}.
\ee

Now let us discuss the power spectrum evaluated at the end of inflation.
The comoving curvature at the attractor stage is given by
\be
{\cal R} = \psi_2+ \frac{H}{\hat{\varphi}_0}\chi_{\rm ATT}
\approx -\frac{1-\kappa m^2}{2} \varphi_i\chi_{\rm ATT},
\ee
where we used the approximations,
$H\approx -m\varphi_i/\sqrt{6}$
and $\psi_2 \approx \kappa m^2\varphi_i\chi_{\rm ATT}/2$
obtained in Ref. \cite{Cho:2014jta} for $\psi_2$ in Eq. \eqref{psi2_XY}.
With $\chi_{\rm ATT} = Q_{\rm ATT}/\omega_{\rm ATT}$
where $\omega_{\rm ATT}^4 \approx (1-4\kappa m^2/3)a^4$
and $a(\tau) = a_i(\tau_i-\tau_0)/(\tau-\tau_0)$
at the attractor stage obtained in Ref. \cite{Cho:2014jta},
the power spectrum becomes
\begin{align}
P_{\cal R} &= \frac{k^3}{2 \pi^2}{\cal R}^2
\approx \frac{k^3}{8\pi^2} (1-\kappa m^2)^2
\varphi_i^2 \left| \frac{Q_{\rm ATT}}{\omega_{\rm ATT}} \right|^2\\
&\approx \frac{2}{\pi}\left( c^2+R^2 + \frac{\pi^2}{16c^2} \right)
\times \frac{(1-\kappa m^2)^2}{(1-4\kappa m^2/3)^{1/2}}
\times \frac{m^2\varphi_i^4}{96\pi^2} \\
&\equiv D_k \times E_\kappa^{\rm S} \times P_{\cal R}^{\rm GR} \label{EkS}\\
&\equiv D_k \times P_{\cal R}^{\rm ATT}.
\end{align}
Here, $P_{\cal R}^{\rm GR} \equiv m^2\varphi_i^4/96\pi^2$ is the power spectrum in GR,
and $E_\kappa^{\rm S} \equiv (1-\kappa m^2)^2/(1-4\kappa m^2/3)^{1/2}$ is the EiBI correction
which is the same as that for the perturbation produced at the attractor stage
obtained in Ref. \cite{Cho:2014jta},
$P_{\cal R}^{\rm ATT} = E_\kappa^{\rm S} P_{\cal R}^{\rm GR}$.
The coefficient $D_k \equiv (2/\pi) (c^2+R^2 + \pi^2/16c^2)$
is the $k$-dependence factor which is the same form
obtained for the tensor perturbation in Ref. \cite{Cho:2014ija}.
As seen in Fig.~1,
$D_k$ exhibits a peculiar rise at low $k$,
while it becomes unity, $D_k \to 1$, at high $k$.
As a result, compared with the power spectrum for the scalar perturbation produced
at the attractor stage,
$P_{\cal R}$ for the perturbation produced at the near-MPS stage is
the same for high $k$-modes,
but exhibits a peculiar peak for low $k$-modes.

The tensor power spectrum in EiBI gravity was obtained in Ref. \cite{Cho:2014ija} as
\be
P_{\rm T} \approx D_k \times E_\kappa^{\rm T}
\times P_{\rm T}^{\rm GR},
\quad\mbox{where}\quad
E_\kappa^{\rm T} = \frac{1}{1+\kappa m^2\varphi_i^2/2}. \label{EkT}
\ee
Here, $P_{\rm T}^{\rm GR}$ is the spectrum for the chaotic inflation model in GR,
and $E_\kappa^{\rm T}$ is the EiBI correction.
Since the factor $D_k$ is common both for the tensor and the scalar perturbations,
the tensor-to-scalar ratio for the perturbation produced at the near-MPS stage is given by
\be
r= \frac{P_{\rm T}}{P_{\cal R}}
\approx \frac{E_\kappa^{\rm T} \times P_{\rm T}^{\rm GR}}{E_\kappa^{\rm S} \times P_{\cal R}^{\rm GR}}
= \frac{(1-4\kappa m^2/3)^{1/2}}{(1-\kappa m^2)^2 (1+\kappa m^2\varphi_i^2/2)} \; r^{\rm GR}
\approx \frac{1+4\kappa m^2/3}{1+\kappa m^2\varphi_i^2/2} \; r^{\rm GR},
\ee
which is exactly the same with that for the perturbations produced at the attractor stage
obtained in Ref. \cite{Cho:2014jta}.
Here, $r^{\rm GR} \sim 0.131$ for $60$ $e$-foldings.
As $\varphi_i \sim {\cal O}(10)$,
the EiBI correction of the tensor spectrum is dominant
and the value of $r$ is lowered.
While the scalar perturbation has been investigated in the limit of $\kappa m^2 \ll 1$,
the result of the tensor perturbation investigated in Ref. \cite{Cho:2014ija}
does not restrict the value of $\kappa m^2\varphi_i^2$ much.
Therefore, within the accuracy of this work, $\kappa m^2 \lesssim {\cal O}(10^{-2})$,
and with $\varphi_i \sim {\cal O}(10)$,
the EiBI correction can be considerably large
and the value of $r$ can be significantly suppressed.

Although the EiBI correction to the tensor-to-scalar ratio can be large,
the correction to the power spectrum itself is tiny.
The EiBI corrections to each power spectrum $P_{\cal R}$ and $P_{\rm T}$
lower their values
because $E_\kappa^{\rm S}<1$ and $E_\kappa^{\rm T}<1$ from
Eqs. \eqref{EkS} and \eqref{EkT}.
However, it is still within the observational bound for the total power spectrum
because $\kappa m^2 \ll 1$ and $P_{\rm T} \ll P_{\cal R}$.

The effect of the peculiar rise by the factor $D_k$
for low $k$-modes is canceled in the tensor-to-scalar ratio.
However, it may appear in observational data for the power spectrum.
It will be observable in the current data
only when the early stage of the inflationary $e$-folding
took place at the near-MPS stage.
Otherwise, the rise corresponds to the very long-wavelength modes
which are not observable today.

\begin{figure}[btph]
\begin{center}
\includegraphics[width=.5\linewidth,origin=tl]{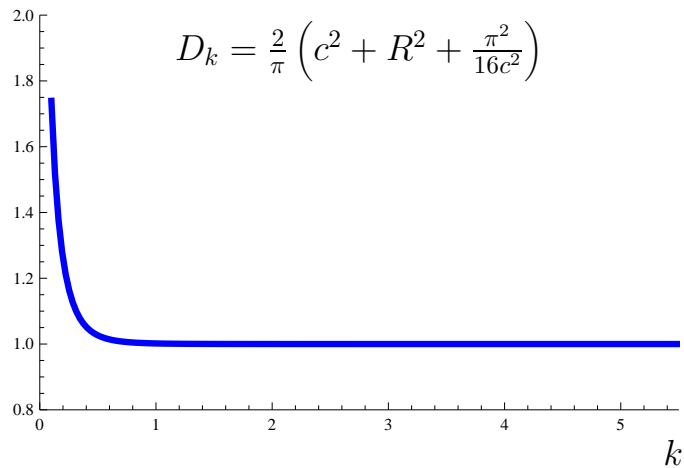}
\end{center}
\caption{Numerical plot of $D_k$ for $\tau_*=1$.
$D_k$ exhibits a peculiar rise for $k\tau_* <1$,
while it approaches unity for $k\tau_* >1$
}
\label{FIG1}
\end{figure}

\section{Conclusions}
In this paper, we investigated the scalar perturbation produced
at the near-MPS (maximal pressure state) which arises at the early stage
of the Universe driven by a massive scalar field in Eddington-inspired Born-Infeld gravity.
This work is a completion of studying the scalar perturbation in this model,
combined with the work \cite{Cho:2014jta} for the scalar perturbation produced at the inflationary attractor stage.

The scalar perturbation pattern is very similar to that \cite{Cho:2014jta} at the attractor stage for high $k$-modes
which mimics the standard chaotic inflation in GR, with a small EiBI correction.
The pattern for low $k$-modes is very similar to that \cite{Cho:2014ija} of the tensor perturbation.
The power spectrum exhibits a peculiar rise,
which can distinguish this EiBI model from the standard chaotic inflation model in GR.
It also possesses the same EiBI correction as for the high $k$-modes.

Because of the similarity, we could buy the results of Refs. \cite{Cho:2014ija} and \cite{Cho:2014jta}.
The tensor-to-scalar ratio $r$ was obtained for all $k$-modes.
The result was the same with that obtained at the attractor stage.
The scalar perturbation increases the value of $r$,
while the tensor spectrum decreases it.
The tensor contribution is dominant,
so $r$ can be lowered significantly.
Recent observational results of BICEP2 \cite{Ade:2014xna} and PLANCK \cite{Ade:2015tva,Adam:2015rua}
for the tensor perturbation have attracted much attention, and are still in dispute.
The direction is that the value of $r$ is to be lowered from the one that BICEP2 observed
(the analyses predict $r_{0.05} < 0.12$ in Ref.~\cite{Ade:2015tva},
and $r_{0.002} < 0.09$ in Ref.~\cite{Adam:2015rua}).
In this sense, our result is very affirmative
since it can decrease $r$ to a very low value.

In this work, we considered the perturbation in the limit of $\kappa m^2 \ll 1$.
From the star formation study in Refs. \cite{Avelino:2012ge,Pani:2011mg,Pani:2012qb},
the theory parameter is very mildly constrained as
$\kappa < 10^{-2} m^5 kg^{-1}s^{-2} \sim 10^{77}$ in Planck unit.
Considering the consistency of our model with inflation,
the parameter is constrained more strongly as
$\kappa \ll m^{-2} \sim 10^{10}$.

After inflation ends, the Universe settles down to the radiation-dominated era
followed by the matter-dominated era.
At these stages after inflation, the Universe is in a very low-energy state
of which evolution is very similar to that in GR as it was investigated in Ref. \cite{Cho:2012vg}.
Therefore, the post-inflationary evolution of perturbations
must be very similar to that in GR.

As we observed in this work, EiBI gravity provides an inflation model in a promising direction.
Other than inflation, EiBI gravity has presented interesting cosmological and astrophysical
results investigated in Refs. \cite{Cho:2012vg,Pani:2011mg,Pani:2012qb,DeFelice:2012hq,Avelino:2012ge,Avelino:2012qe,Casanellas:2011kf,
Liu:2012rc,Delsate:2012ky,Pani:2012qd,Cho:2013usa,Scargill:2012kg,Kim:2013nna,Kim:2013noa,Du:2014jka,Ji:2014hna,
Fu:2014raa,Harko:2014nya}.
In the gravity theory point of view, the unitarity problem related with
the ghost graviton mode of the theory is an important issue.
The existence of the ghost in EiBI theory is not clear yet.
Very recently in Ref. \cite{Schmidt-May:2014tpa}, the authors investigated the ghost problem in bimetric theory,
including EiBI gravity as an example.
According to their work, when matter is absent,
there is no ghost since EiBI gravity is equivalent to GR.
When matter is coupled to EiBI gravity, however, it is not entirely clear
whether or not, the theory suffers from the ghost instability.
In our investigation, we introduced eight gravitational perturbation fields
for two metrics and one matter perturbation field.
Eight gravitational perturbation fields can be solved
in terms of the matter perturbation field $\chi$
with our choice of the gauge conditions.
All the solutions for these fields look regular.
However, we cannot conclude that there is no ghost.
Even when there is a ghost, the related physical quantities can behave regularly
depending on the coupling to others.
As a whole, the ghost problem in EiBI gravity needs to be investigated
in a rigorous way in the action level
in the presence of the matter coupling.
We hope that this is resolved in the future.

\section*{Acknowledgement}
The authors are grateful to Jihad Mourad, Daniele Steer, Mikael von Strauss,
and Jinn-Ouk Gong for very helpful discussions.
This work was supported by the grant from the National Research Foundation
funded by the Korean government, No. NRF-2012R1A1A2006136.

 \end{document}